\documentclass[aps,pra,twocolumn,showpacs,floatfix]{revtex4}

\usepackage{amsmath}
\usepackage{amssymb}
\usepackage{graphicx}
\usepackage{epstopdf}
\usepackage{verbatim}
\usepackage{color}
\usepackage{appendix}
\usepackage[colorlinks, linkcolor=blue, citecolor=blue, urlcolor=blue,breaklinks=true]{hyperref}
\usepackage{soul}

\begin{document}

\newcommand{\ie}{{\it i.e.}}

\newcommand{\bquote}{\quotedblbase{}}
\newcommand{\equote}{\textquotedblright{ }}

\newcommand{\cl}[1]{\mathcal{#1}} 
\newcommand{\on}[1]{#1^\dag \, #1} 
\newcommand{\comm}[2]{\left[#1 \, , \, #2 \right]} 
\newcommand{\ev}[1]{ \left \langle #1 \right \rangle } 
\newcommand{\anticomm}[2] { \left \{ #1 \, , \, #2 \right \} } 

\newcommand{\opvec}[1]{\hat{\textbf{#1}}}
\newcommand{\tildevec}[1]{\tilde{\textbf{#1}}}
\newcommand{\boldvec}[1]{\boldsymbol{#1}}

\newcommand{\ket}[1]{\left|{#1}\right\rangle}
\newcommand{\bra}[1]{\left\langle{#1}\right|}

\newcommand{\norm}[1]{\| #1 \|}

\newcommand{\re}{\textnormal{Re}}
\newcommand{\im}{\textnormal{Im}}

\title{Finite-size scaling in the quantum phase transition of the open-system Dicke-model}
\author{G. K\'onya}
\author{D. Nagy}
\author{G. Szirmai}
\author{P. Domokos}

\affiliation{Institute for Solid State Physics and Optics, Wigner Research Centre, Hungarian Academy of Sciences, H-1525 Budapest P.O. Box 49, Hungary}
\begin{abstract}
Laser-driven Bose-Einstein condensate of ultracold atoms loaded into a lossy high-finesse optical resonator exhibits critical behavior and,  in the thermodynamic limit, a phase transition between stationary states of different symmetries. The system realizes an open-system variant of the celebrated  Dicke-model.  We study the transition for a finite number of atoms by means of a Hartree-Fock-Bogoliubov method adapted to a damped-driven open system. The finite-size scaling exponents are determined and a clear distinction between the non-equilibrium and the equilibrium quantum criticality is found.
\end{abstract}


\maketitle

\section{Introduction}

Ultracold gases of atoms manipulated by magnetic and laser fields proved to be a powerful tool to study quantum many-body problems \cite{Bloch2008Manybody,Giorgini2008Theory}. Paradigmatic models, such as the Hubbard model, which have been introduced to phenomenologically describe fundamental effects of solid-state physics, can now be realized experimentally with ultracold atoms \cite{Jaksch2005Cold}. The atom-based implementation also offers the intriguing possibility of tuning the system parameters and, thereby, reaching so-far unexplored regimes of these models. Thus, beyond quantum simulation of various many-body problems \cite{Greiner2002Quantum}, the high-degree of control of the dynamics in a reduced Hilbert-space can lead to novel applications in quantum measurement and also in quantum information processing.  

Experimental setups of cavity quantum electrodynamics (cavity QED) allow for the realization of quantum gases with long-range interaction \cite{Colombe2007Strong,Brennecke2007Cavity,Slama2007Superradiant,Murch2008Observation,Wolke2012Cavity}. Instead of short-range collisions, the dominant atom-atom interaction is mediated by the electromagnetic radiation field enclosed into a high-finesse resonator \cite{Munstermann2000Observation,Asboth2004Correlated}. The many-body ensemble couples globally to the resonator field, which situation, in certain cases, can be described by collective spin-boson models. In particular, there has been recently a breakthrough in the realization of the Dicke model \cite{Dicke1954Coherence}. It is a paradigmatic model of quantum optics describing the interaction of a single mode of the radiation field with a collection of two-level atoms (spins),
\begin{equation}
\label{eq:DickeHamiltonian}
\hat{H} =  \omega_C \; \on{\hat{a}} \; + \;  \omega_R \; \hat{S}_z \; + \; \frac{y}{\sqrt{N}} \; (\hat{a}^\dagger + \hat{a} ) \; \hat{S}_x \; ,
\end{equation}
where $\hat{a}$ is the bosonic annihilation operator of the field mode, and the collective spin $\opvec{S} = \sum_{i=1}^N \hat{\boldsymbol{\sigma}}^{(i)}$, the index $i$ labeling the spin $\hat{\boldsymbol{\sigma}}$ associated with the individual two-level atoms.   The main interest in this model is related to the predicted quantum phase transition \cite{Hepp1973Equilibrium,Hillery1985Semiclassical,Emary2003Chaos} between the normal phase (all atom is in the ground state, and the radiation field is vacuum) and a superradiant phase  (the atoms and the mode have a coherent polarization $\ev{\opvec{S}} \neq 0$ and field amplitude $\ev{\hat{a}} \neq 0$, respectively) at the coupling strength $y= \sqrt{\omega_C \, \omega_R}$, \ie, when it reaches the geometric mean of the eigenfrequencies $\omega_C$ and $\omega_R$ of the field mode and the atoms, respectively. However, it is generally believed  impossible to achieve such a strong dipole coupling between atoms and the radiation field  in practice.

\begin{figure}[ht!]
\centering
\includegraphics[width=\columnwidth]{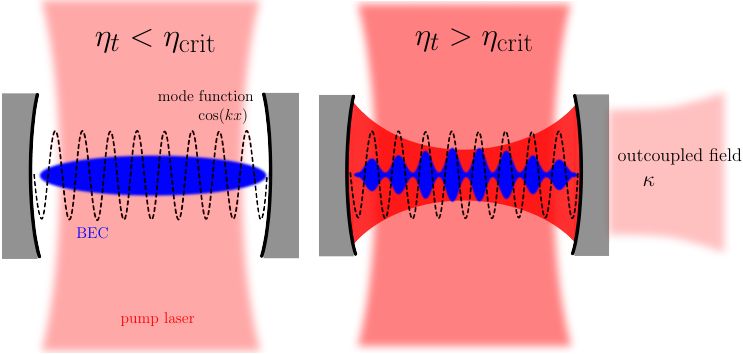}
\caption{(Color online) Self-organization phase transition of a BEC in a cavity. Below a threshold in the transverse driving field (left) the condensate is quasi-homogeneous, and there is no radiation field inside the cavity. Above threshold (right), a standing matter wave of period $\lambda$ appears that scatter photons into the cavity.}
\label{fig:scheme}
\end{figure}

A completely different approach, based on the motional excitations of a Bose--Einstein condensate loaded into an optical cavity, led to the first observation of the phase transition in the Dicke model \cite{Baumann2010Dicke}. The Hamiltonian describing the coherent coupling of a single mode of the resonator with two momentum eigenstates of a degenerate quantum gas driven by a laser perpendicular to the cavity axis (see Fig.~\ref{fig:scheme}) was shown to be formally equivalent to the Dicke model \cite{Nagy2010DickeModel,Baumann2010Dicke}. Beyond the mapping of the phase diagram  \cite{Baumann2010Dicke}, the spontaneous symmetry breaking \cite{Baumann2011Exploring} as well as the mode softening in the excitations spectrum at the critical point \cite{Mottl2012RotonType} have been demonstrated experimentally. This approach does not involve electronic states of the atoms other than the ground state. The characteristic frequency scale of the interaction is determined by the recoil frequency $\omega_R = k^2/2m$ of the specific atom ($k$ is the wavenumber of the driving laser, $m$ is the atomic mass), which is typically in the kHz range, well below the internal, electronic energy scale.  The coupling strength can be tuned continuously by adjusting the power of the driving laser field.  The Dicke model phase transition with ultracold atoms is, in fact, the zero temperature limit of the atomic self-organization in a cavity, which has been predicted in \cite{Domokos2002Collective} and demonstrated in \cite{Black2003Observation} for cold but not ultracold atoms.

The cavity-based realization is an open system. The atoms are driven by a monochromatic laser source, scatter photons into the cavity which then leak out through the end mirrors.  Therefore the Dicke-type Hamiltonian does not provide for a full description of the system. The critical behavior has been reinvestigated recently for the stationary state of the driven and damped system. Using a generalized Bogoliubov theory to treat the thermodynamic limit, it was shown that the non-equilibrium system also  exhibits a dynamical quantum ($T=0$) phase transition  \cite{Nagy2011Critical}. The critical point as well as the mean field solution are only slightly modified with respect to the phase transition in the ground state of \eqref{eq:DickeHamiltonian}. However, the correlation functions describing the quantum fluctuations differ significantly in the two,  equilibrium and non-equilibrium, cases. This difference can be captured, for example, by looking at the critical exponents of the diverging photon number close to the critical point.  There is also a drastic modification of the atom-field entanglement, since the singularity of the logarithmic negativity of the ground state \cite{Lambert2004Entanglement} at the critical point is regularized, though some finite entanglement is still found in the steady state of the driven open system. 

In this paper we study the finite-size scaling in the phase transition of the open-system Dicke model.  On the one hand, as the experiments involve obviously a finite number of atoms, it is necessary to calculate the  correlation functions for a finite system. On the other hand, theoretically, the finite-size scaling gives insight into the criticality and its classification. Various systems can show similar critical behavior, that leads to the concept of universality: the systems sharing the same critical exponents correspond to the same universality class. For the ground state of the Dicke model the finite-size corrections and scaling exponents are known \cite{Vidal2006Finitesize}, and show a good agreement with the critical scaling behavior of infinitely coordinated spin systems described by the Lipkin--Meskhov--Glick model \cite{Dusuel2005Continuous}. This correspondence has been investigated both analytically \cite{Liberti2010Finitesize} and by numerical calculations \cite{Reslen2005Direct}. It is well known that for infinitely coordinated systems the mean-field approach is exact in the thermodynamic limit and the large-size critical behavior is related to the upper critical dimensionality of the corresponding short range system \cite{Botet1983Largesize}. Note that  the short-range equivalent of the Dicke model is unknown. Here we will consider exponents for a non-equilibrium quantum critical system, for the open Dicke model, in order to make a first step towards the classification. With the help of a scaling hypothesis consistent with the numerical findings we identify the two critical exponents describing the universal behavior of the system.

The mean-field theory we used previously is suitable to describe the thermodynamic limit \cite{Nagy2011Critical}. In this paper we develop a Hartree--Fock--Bogoliubov (HFB) theory for the finite open system. The population of the excitation modes above the condensate is not due to a finite temperature but the atom-light interaction leads to a quantum depletion of the ground state.  Close to the critical point, a macroscopic part of the atoms can be out of the condensate. If the total number of atoms is finite, the quantum fluctuations represented by such a macroscopic depletion are expected to yield a considerable back action on the mean field. This effect is taken into account by the HFB theory.

The structure of this paper is as follows. In Sec.~\ref{sec:Description} we present the system of ultracold atoms coupled to a cavity field together with the physical approximations adopted in the description, and introduce the basic equations of motion. In Sec.~\ref{sec:HartreeFockBogoliubov} we derive the HFB approximation generalized to an open system. The theory involves a self-consistency loop, therefore, a better understanding of the approach can be gained by the details of the iterative algorithm of the solution which we summarize in a separate subsection. The results are presented then in Sec.~\ref{sec:Results}. The phase transition is exhibited, however, we  focus mostly on the finite-size scaling of the quantum criticality.  We conclude and give an outlook in Sec.~\ref{sec:Conclusion}. Some lengthy expressions are showed in Appendices.

\section{Description of the system} 
\label{sec:Description}

We study the quantum phase transition of a Bose--Einstein condensate which is placed into an optical resonator and pumped by a laser from a  direction perpendicular to the resonator axis \cite{Nagy2008Selforganization,Keeling2010Collective,Gopalakrishnan2009Emergent,Vidal2010Quantum}. We consider a single cavity mode which is quasi-resonant with the laser pump frequency $\omega$. The detuning between the laser and the atom, defined as $\Delta_A=\omega-\omega_A$, is large enough so that the internal dynamics of the atoms can be adiabatically eliminated and only the center-of-mass motion takes part in the dynamics.  For simplicity, the motion is considered in one dimension along the cavity axis direction denoted by $x$. The cavity length is $L$ and the mode function in the relevant direction is $f(x)=\cos(kx)$. The atoms are represented by the second-quantized boson field operator $\hat{\Psi} (x)$, whereas the resonator mode is described by the operator $\hat{a}$.

The Hamiltonian in units of $\hbar=1$ is
\begin{multline} \label{H_beccav}
\hat{H} = \omega_{C} \; \on{\hat{a}}
 + \int_{-\frac{L}{2}}^{+\frac{L}{2}} \hat{\Psi}^{\dagger}(x)
\Biggl( -\frac{1}{2m} \frac{d^{2}}{d x^{2}}   \\
+U_{0} \; \on{\hat{a}} \, \cos^{2} (kx) \\
+ \eta_t \, \cos(kx) \, \left( \hat{a}^\dag \, e^{-i \omega t} + \hat{a} \, e^{i \omega t} \right) \Biggr) \hat{\Psi}(x) \, dx\,.
\end{multline}
The first term gives the energy of cavity photons of frequency $\omega_C$, the second one the motional energy of the atom field. This latter is characterized by the so-called recoil frequency $\omega_R=k^2/2m$. The dispersive atom-light interaction yields two terms. Firstly, there is an atomic refractive index term with coupling strength  $U_0 = \frac{g^2}{\Delta_A}$ where $g$ is the single-photon Rabi frequency of the cavity mode. Secondly,  there appears an effective cavity driving by means of Raman scattering from the transverse pump laser with amplitude $\eta_t \; = \; \frac{\Omega_{L} \, g} {\Delta_A}$, where $\Omega_{L}$ is the laser Rabi frequency. In this term the time evolution at the pump frequency $\omega$ is made explicit and we used the rotating wave approximation.  We neglected the atom-atom s-wave collision term, because the optical interaction dominates in the experiments.

The system is spatially periodic with respect to the cavity mode wavelength, therefore a discrete Fourier basis is convenient to decompose the atom field, 
\begin{equation} \label{Psi_Fourier_decomposition}
\hat{\Psi} (x)= \sqrt{\frac{1}{L}} \, \hat{c}_0 + \sum_{n=1}^{\infty} \sqrt{\frac{2}{L}} \, \cos(nkx) \, \hat{c}_n  \,.
\end{equation}
Since parity is conserved in the system, no transition from the even cosine functions to the odd sine functions is allowed. Assuming initially a homogeneous condensate, we can keep only the subspace of the cosine modes.  Let us introduce the notation $\opvec{c}= \left(\hat{c}_0 \, , \, \hat{c}_1 \, , \, \hat{c}_2 \, , \, ... \right)^T$ for a column, and
$\opvec{c}^\dag= \left(\hat{c}_0^\dag \, , \, \hat{c}_1^\dag \, , \, \hat{c}_2^\dag \, , \, ... \right)$ for a row vector to make the forthcoming expressions compact. For example, the total number of atoms reads  $\hat{N}=\opvec{c}^\dag \, \opvec{c}$ and $N=\ev{\hat{N}}$.  
Obviously, in the numerical calculation,  the Fourier expansion \eqref{Psi_Fourier_decomposition} is truncated at a cutoff index $n_{\rm max}$. 

The Hamiltonian \eqref{H_beccav} conserves the atom number, i.e., $\hat{H}$ and $\hat{N}$ commute, which enables us to introduce the grand canonical Hamiltonian given by $\hat{K}=\hat{H}-\mu \hat{N}$, where $\mu$ is the chemical potential. The grand canonical Hamiltonian can be built up from quadratic forms, and reads in a frame rotating at the laser frequency $\omega$
\begin{subequations}
\begin{multline} \label{K_beccav_matrix}
\hat{K}= 
-\Delta_{C} \, \on{\hat{a}} + \omega_R \, \left( \opvec{c}^\dag \, \textbf{M}^{(0)} \, \opvec{c} \right) \\
+ \frac{\sqrt{2}}{2} \eta_t \left( \hat{a}^\dag + \hat{a} \right) \left( \opvec{c}^\dag \, \textbf{M}^{(1)} \, \opvec{c} \right) \\
+\frac{1}{4} \, U_0 \, \on{\hat{a}} \, \left( \opvec{c}^\dag \, \textbf{M}^{(2)} \, \opvec{c} \right)
-\mu \, \opvec{c}^\dag \, \opvec{c} \, ,
\end{multline}
where $\Delta_C=\omega-\omega_C$. The kernel matrices $\textbf{M}^{(j)}$ are given by:

\begin{equation}
\textbf{M}^{(0)} = 
\left( {\begin{array}{*{20}{c}}
   {{0^2}} & {} & {} & {} & {} & {}  \\
   {} & {{1^2}} & {} & {} & {} & {}  \\
   {} & {} & {{2^2}} & {} & {} & {}  \\
   {} & {} & {} & {{3^2}} & {} & {}  \\
   {} & {} & {} & {} & \cdot & {}  \\
   {} & {} & {} & {} & {} & \cdot  \\
\end{array}} \right)
\end{equation}

\begin{equation}
\textbf{M}^{(1)} = 
\left( {\begin{array}{*{20}{c}}
   0 & 1 & {} & {} & {} & {}  \\
   1 & 0 & {\frac{1}{{\sqrt 2 }}} & {} & {} & {}  \\
   {} & {\frac{1}{{\sqrt 2 }}} & 0 & {\frac{1}{{\sqrt 2 }}} & {} & {}  \\
   {} & {} & {\frac{1}{{\sqrt 2 }}} & 0 & {\frac{1}{{\sqrt 2 }}} & {}  \\
   {} & {} & {} & {\frac{1}{{\sqrt 2 }}} & 0 & \cdot  \\
   {} & {} & {} & {} & \cdot & \cdot  \\
\end{array}} \right)
\end{equation}

\begin{equation}
\textbf{M}^{(2)} = 
\left( {\begin{array}{*{20}{c}}
   2 & 0 & {\sqrt 2 } & {} & {} & {}  \\
   0 & 3 & 0 & 1 & {} & {}  \\
   {\sqrt 2 } & 0 & 2 & 0 & 1 & {}  \\
   {} & 1 & 0 & 2 & 0 & \cdot  \\
   {} & {} & 1 & 0 & \cdot & \cdot  \\
   {} & {} & {} & \cdot & \cdot & \cdot  \\
\end{array}} \right)
\end{equation}
\end{subequations}
Note that the matrix $\textbf{M}^{(j)}$ is a real, symmetric band matrix and the left and right bandwidths are given by the index $j$. This means that the laser driving couples only adjacent modes, while the light shift term couples second neighbors.

The open system description, taking into account the irreversible loss of photons through the out-coupling mirror, relies on the Heisenberg--Langevin equations of motion \cite{CohenTannoudjiBook}
\begin{subequations}
 \label{eq:EquationsOfMotion}
\begin{gather}
i \frac{d}{dt} \hat{a} (t) = \comm{\hat{a} (t)}{\hat{K}} - i \kappa \, \hat{a} (t) + i \hat{\xi} (t) \\
i \frac{d}{dt} \opvec{c} (t) = \comm{\opvec{c} (t)}{\hat{K}} \; \; ,
\end{gather}
\end{subequations}
where the cavity photon loss rate is $2\kappa$, and $\hat{\xi} (t)$ is a white noise. The noise operator has zero mean value, and it's only non-vanishing second-order correlation function is 
\begin{equation}
\ev{ \hat{\xi} (t) \, \hat{\xi}^\dag (t')} = 2 \, \kappa \, \delta ( t -t' ) \; .
\end{equation}

We use mean-field approach to describe the system. The boson operators of the photon and atomic modes are split into mean-field amplitudes and fluctuations,
\begin{subequations} \label{a_c_mean_field_decomposition}
\begin{gather}
\hat{a}= \sqrt{N_c} \, \alpha \,+ \, \tilde{a} \\
\opvec{c}= \sqrt{N_c} \, \boldvec{\gamma} \, + \, \tildevec{c} \; .
\end{gather}
\end{subequations}
where $\; \ev{\tilde{a}}=0 \;$ and $\; \ev{\tildevec{c}}=\textbf{0} \;$ by definition. $\alpha$ is the coherent cavity field and $\boldvec{\gamma}$ is the condensate wave-function in the basis given by \eqref{Psi_Fourier_decomposition}. This latter vector is normalized as $\boldvec{\gamma}^\dag \, \cdot \, \boldvec{\gamma} = 1$, thereby $N_c$ is the number of condensed atoms. We assume that the value of $N_c$ is a constant of time \footnote{This assumption will only be required later, when we consider the steady state of the system, but it simplifies the calculation without losing interesting effects.}. Finally, the operators $\tilde{a}$ and $\tildevec{c}$ account for the quantum fluctuations around the coherent mean value, e.g. for the atoms outside the condensate. 

The coupling constants in the interaction terms can be redefined as
\begin{subequations} \label{y,u_def}
\begin{gather}
y=\sqrt{2 \, N_c} \; \eta_t \\
u= \frac{1}{4} \; N_c \; U_0  \; ,
\end{gather}
\end{subequations}
The new parameters $u$ and $y$ have a constant value in the thermodynamic limit which is defined by $N \rightarrow \infty$ and $L\rightarrow \infty$, while keeping the density $N/L$ constant. This follows from the fact that $g \sim 1/\sqrt{L}$, which leads to $\eta_t \sim 1/\sqrt{L}$ and $U_0 \sim 1/L$.

On substituting the mean-field decomposition into the equations of motion \eqref{eq:EquationsOfMotion}, a hierarchy of terms according to the different  orders of the $\tilde{a}$ and $\tildevec{c}$ operators can be established. As this intermediate step is needed only for the derivation, we display the lengthy expressions in Appendix \ref{Heisenberg_mean-field_dec}, in Eqs.~\eqref{alpha_tilde{a}_eq_of_motion} and \eqref{gamma_tilde{c}_eq_of_motion}. These include the equations of motion both for the mean fields and for the fluctuation operators. To make the set of equations complete, one needs the relation  
\begin{equation}
N_c = N - \ev{\tildevec{c}^\dag \, \tildevec{c}}\,,
\end{equation}
for the number of condensed atoms, which follows from the mean-field decomposition and from the normalization of $\boldvec{\gamma}$. Without further approximations, the set of equations involve nonlinear, operator valued differential equations, which cannot be solved in practice.

\section{Hartree-Fock-Bogoliubov theory}
\label{sec:HartreeFockBogoliubov}

In this section we present how we go beyond the Bogoliubov-type mean field analysis that we used previously to describe the damped-driven system of a cavity mode coupled to motional excitations of a BEC \cite{Nagy2011Critical}. The method is a kind of Hartree-Fock-Bogoliubov (HFB) approximation \cite{Griffin1996Conserving} generalized to the steady-state of an open system. 
It is needed although the system is at $T=0$, because the values of the second order correlation functions are large near the phase transition point. The higher-order terms, neglected in the bare Bogoliubov-approximation,  can have a considerable effect on the mean value equations. 
Moreover, the HFB approach gives atom number dependent corrections to the results obtained for the thermodynamic limit, which we will use for the finite-size scaling.

\subsection{The self-consistent mean field equations}

The starting point of the theory is the set of Eqs.~\eqref{alpha_tilde{a}_eq_of_motion} and \eqref{gamma_tilde{c}_eq_of_motion}. In the simple Bogoliubov-approximation, the terms of second and of third order in the fluctuation operators are simply neglected, so we obviously get linear equations for the fluctuation operators. At variance, in the HFB method, these higher order terms are approximated by using lower order terms as shown in Appendix \ref{Heisenberg_mean-field_dec}. The second order terms are substituted by their expectation values. In the third order terms, we substitute the product of two operators by its expectation value, and leave the third one as an operator. There are three possible ways of doing this, so we do it in all possible ways and each third order term is approximated by the sum of three linear terms. The equations of motion for the fluctuation operators  become then linear in the HFB approximation.  

The HFB equation of motion for the cavity mean field is
\begin{subequations}
 \label{eq:HFBmean}
\begin{multline}
i \frac{d}{dt} \alpha = ( \Omega - i\kappa ) \alpha + \frac{1}{2} y \, \left( \boldvec{\gamma}^\dag \textbf{M}^{(1)} \boldvec{\gamma} + \frac{1}{N_c} \, \ev{\tildevec{c}^\dag \textbf{M}^{(1)} \tildevec{c}} \right) \\
+ \frac{u}{N_c} \left( \ev{\tildevec{c}^\dag \, \tilde{a}} \textbf{M}^{(2)} \boldvec{\gamma} + \boldvec{\gamma}^\dag \textbf{M}^{(2)} \ev{\tildevec{c} \, \tilde{a} }  \right) \;,
\end{multline}
and for the condensate it reads
\begin{multline}
i \frac{d}{dt} \boldvec{\gamma} = \left( \textbf{M} - \mu \textbf{I} \right) \boldvec{\gamma} + \frac{1}{N_c} \; \Biggl[
\; \frac{1}{2} y \; \textbf{M}^{(1)} \left( \ev{\tilde{a}^\dag \, \tildevec{c}} + \ev{\tilde{a} \, \tildevec{c}} \right) \\
+ u \left( \alpha^* \textbf{M}^{(2)} \ev{\tilde{a} \, \tildevec{c}}  + \alpha \textbf{M}^{(2)} \ev{\tilde{a}^\dag \, \tildevec{c}} \right)\; \Biggr] \; .
\end{multline}
\end{subequations}
Here we defined the renormalized cavity frequency
\begin{equation} \label{Omega_def_old_basis}
\Omega = -\Delta_C + u \; \boldvec{\gamma}^\dag \textbf{M}^{(2)} \boldvec{\gamma} + \frac{u}{N_c} \; \ev{\tildevec{c}^\dag \textbf{M}^{(2)} \tildevec{c}} \,,
\end{equation}
which includes the resonance shift due to the dispersive interaction with the atoms. We also introduce the real and symmetric matrix
\begin{multline} \label{M_def}
\textbf{M}= \omega_R \; \textbf{M}^{(0)} +\frac{1}{2} y \, \left( \alpha^* + \alpha \right) \, \textbf{M}^{(1)}  \\
 + u \; \alpha^* \alpha \; \textbf{M}^{(2)} + \frac{u}{N_c} \, \ev{\on{\tilde{a}}} \, \textbf{M}^{(2)} \,,
\end{multline}
which includes the eigenenergy and the coupling between the atomic modes via the cavity field. This coupling will be diagonalized in the next subsection.

The equation of motion for the fluctuating term of the cavity field is
\begin{subequations}
 \label{eq:HFBfluct}
\begin{multline} \label{a_eq_old_basis}
i \frac{d}{dt} \tilde{a} = \left( \Omega - i \kappa \right) \tilde{a} + u \alpha \left( \boldvec{\gamma}^\dag \textbf{M}^{(2)} \tildevec{c} +
\tildevec{c}^\dag \textbf{M}^{(2)} \boldvec{\gamma} \right) \\
+ \frac{1}{2} y \, \left( \boldvec{\gamma}^\dag \textbf{M}^{(1)} \tildevec{c} + \tildevec{c}^\dag \textbf{M}^{(1)} \boldvec{\gamma} \right)  \\
+ \frac{u}{N_c} \left( \ev{\tildevec{c}^\dag \, \tilde{a}} \textbf{M}^{(2)} \tildevec{c} + \tildevec{c}^\dag \textbf{M}^{(2)} \ev{\tildevec{c} \, \tilde{a}} \right) + i \hat{\xi} (t) \; .
\end{multline}
The dynamics of the atomic excitation modes above the condensate is 
\begin {multline} \label{c_eq_old_basis}
i \frac{d}{dt} \tildevec{c} = \left( \textbf{M} -\mu \textbf{I} \right) \tildevec{c} + \frac{1}{2} y  \left( \tilde{a}^\dag + \tilde{a} \right)
\textbf{M}^{(1)} \boldvec{\gamma} \\
+u \left( \alpha^* \tilde{a} + \tilde{a}^\dag \alpha \right) \textbf{M}^{(2)} \boldvec{\gamma} \\
 + \frac{u}{N_c} \left[ \tilde{a}^\dag \, \textbf{M}^{(2)}
\ev{ \tilde{a} \, \tildevec{c} } + \tilde{a} \, \textbf{M}^{(2)} \ev{\tilde{a}^\dag \tildevec{c}} \right] \,.
\end{multline}
\end{subequations}

Equations~\eqref{eq:HFBmean} and \eqref{eq:HFBfluct} fully define the HFB theory. The terms proportional to $1/N_c$ are new with respect to the Bogoliubov-approach. They express the back-action of the fluctuations on the mean values. Therefore the mean field and the correlation functions have to be determined self-consistently.  We will look for the steady state solution of the coupled mean-field and fluctuation equations. This means that the mean fields $\alpha$ and $\boldvec{\gamma}$ as well as the second order correlation functions built from $\tilde{a} (t)$ and $\tildevec{c} (t)$ have a constant value in time. As we noted earlier, it would be enough to set the condensate number $N_c$ a constant of time at this point.  The fluctuation operators, however, still strongly fluctuate in time because of the quantum noise $\hat{\xi} (t)$ driving terms.

\subsection{Decoupling the atomic modes}

In the first step, we resolve the direct coupling between the atomic modes $\tildevec{c}$  as described in Eq.~\eqref{c_eq_old_basis} by the matrix $\textbf{M}$. Since $\textbf{M}$ is real and symmetric, it can be diagonalized by an orthogonal transformation $\textbf{O}$, where $\textbf{O} \, \textbf{O}^T = \textbf{O}^T \, \textbf{O} =\textbf{I}$. Let us perform the transformation on the atomic modes $\boldvec{\gamma} = \textbf{O} \, \boldvec{\beta}$ and $\tildevec{c} = \textbf{O} \, \tildevec{b}$ and simultaneously
\begin{subequations} 
\label{matrix_orthogonal_transformation}
\begin{gather}
\textbf{M} \; \rightarrow \; \textbf{O}^T \, \textbf{M} \, \textbf{O} = \boldvec{\Lambda}  \label{matrix_orthogonal_transformation_a}
\\
\textbf{M}^{(j)} \; \rightarrow \; \textbf{O}^T \, \textbf{M}^{(j)} \, \textbf{O} = \tildevec{M}^{(j)} \label{matrix_orthogonal_transformation_b}
\; ,
\end{gather}
\end{subequations} 
where the matrix $\boldvec{\Lambda}$ is diagonal by the definition of the orthogonal matrix $\textbf{O}$. Thanks to the orthogonality of the transformation, the vector $\boldvec{\beta}$ is also normalized: $\boldvec{\beta}^\dag \; \boldvec{\beta} = 1$ and the number of condensed atoms is given by 
\begin{equation}
\label{eq:N_BEC}
N_c = N - \ev{\tildevec{b}^\dag \, \tildevec{b}}\,.
\end{equation}

The steady-state mean-field equations in the new basis read
\begin{subequations}
\begin{multline} \label{alfa_mean_field}
0 = ( \Omega - i\kappa ) \alpha + \frac{1}{2} y \, \left( \boldvec{\beta}^\dag \tildevec{M}^{(1)} \boldvec{\beta} + \frac{1}{N_c} \, \ev{\tildevec{b}^\dag \tildevec{M}^{(1)} \tildevec{b}} \right) \\
+ \frac{u}{N_c} \left( \ev{\tildevec{b}^\dag \, \tilde{a}} \tildevec{M}^{(2)} \boldvec{\beta} + \boldvec{\beta}^\dag \tildevec{M}^{(2)} \ev{\tildevec{b} \, \tilde{a} }  \right) \; ,
\end{multline}
\begin{equation} \label{beta_mean_field}
0 = \left( \boldvec{\Lambda} -\mu \textbf{I} \right) \boldvec{\beta} + \textbf{R} \; ,
\end{equation}
\end{subequations}
where $\Omega$ in the transformed basis is
\begin{equation} \label{Omega_def_new_basis}
\Omega = -\Delta_C + u \; \boldvec{\beta}^\dag \tildevec{M}^{(2)} \boldvec{\beta} + \frac{u}{N_c} \ev{ \tildevec{b}^\dag \tildevec{M}^{(2)} \tildevec{b} } \; ,
\end{equation}
 and the new quantity $\textbf{R}$ was introduced:
\begin{multline}
\textbf{R} = \frac{1}{N_c} \Biggl[ \; \frac{1}{2} y \; \tildevec{M}^{(1)} \left( \ev{\tilde{a}^\dag \tildevec{b}} + \ev{\tilde{a} \tildevec{b}} \right) \\
 + u \left( \alpha^* \tildevec{M}^{(2)} \ev{\tilde{a} \tildevec{b}} + \alpha \tildevec{M}^{(2)} \ev{\tilde{a}^\dag \tildevec{b}} \right)  \Biggr] \label{R_def}
\end{multline} 
The value of the chemical potential can be determined  by multiplying both sides of Eq.~\eqref{beta_mean_field} by $\boldvec{\beta}^\dag$ from the left, 
\begin{equation} \label{mu_def}
\mu = \boldvec{\beta}^\dag \, \Lambda \, \boldvec{\beta} + \boldvec{\beta}^\dag \, \textbf{R} \,,
\end{equation}
and we made use of the fact that $\boldvec{\beta}$ is normalized to one. 

\subsection{Linearized equations of fluctuations}

The equations of the fluctuation operators in the new basis are
\begin{subequations}
\label{eq:LinFluctNewBasis}
\begin{multline} \label{a_eq_new_basis}
i \frac{d}{dt} \tilde{a} = \left( \Omega - i \kappa \right) \tilde{a} + u \alpha \left( \boldvec{\beta}^\dag \tildevec{M}^{(2)} \tildevec{b} + \tildevec{b}^\dag \tildevec{M}^{(2)} \boldvec{\beta} \right) \\
 + \frac{1}{2} y \, \left( \boldvec{\beta}^\dag \tildevec{M}^{(1)} \tildevec{b} + \tildevec{b}^\dag \tildevec{M}^{(1)} \boldvec{\beta} \right) \\
 + \frac{u}{N_c} \, \left( \ev{\tildevec{b}^\dag \tilde{a}} \tildevec{M}^{(2)} \tildevec{b} + \tildevec{b}^\dag \tildevec{M}^{(2)} \ev{\tildevec{b} \tilde{a}} \right) + i \hat{\xi} (t) \,,
\end{multline}
\begin{multline} \label{b_eq_new_basis}
i \frac{d}{dt} \tildevec{b} = \left( \boldvec{\Lambda} - \mu \textbf{I} \right) \tildevec{b} + \frac{1}{2} y \left( \tilde{a}^\dag + \tilde{a} \right) \tildevec{M}^{(1)} \boldvec{\beta}  \\
+ u \left( \alpha^* \tilde{a} + \tilde{a}^\dag \alpha \right) \tildevec{M}^{(2)} \boldvec{\beta} \\
 + \frac{u}{N_c} \left[ \tilde{a}^\dag \; \tildevec{M}^{(2)} \ev{\tilde{a} \tildevec{b}} + \tilde{a} \; \tildevec{M}^{(2)} \ev{\hat{a}^\dag \tildevec{b}} \right] \,.
\end{multline}
\end{subequations}
Note that, in the new basis, the $\tildevec{b}$ operators are also decoupled from each other. Let us put the fluctuation operators into vector form, $\tildevec{v} (t) = \left( \tilde{a} (t) , \; \tilde{a}^\dag (t) , \; \tildevec{b} (t) , \; \tildevec{b}^\dag (t) \right)^T $, then the coupled linear equations \eqref{eq:LinFluctNewBasis} can be written in the compact form
\begin{equation} \label{lin_fluctuation_eq}
i \frac{d}{dt} \; \tildevec{v} (t) \; = \; \textbf{F} \; \tildevec{v} (t) \; + \; i \, \opvec{q} (t) \; \; ,
\end{equation}
where the  $\textbf{F}$ matrix contains the coefficients and $\opvec{q}(t)= \left( \hat{\xi}(t) , \; \hat{\xi}^\dag (t) , \; \textbf{0} , \; \textbf{0} \right)^T$ is the noise. The noise correlations are given by
\begin{equation}
\ev{\hat{q}_n (t) \, \hat{q}_j (t')} = D_{\, n \, j} \, \delta \left( t - t' \right) \; \; ,
\end{equation}
where the $\textbf{D}$ diffusion matrix reads 
\begin{equation}
\textbf{D} = 
\left( {\begin{array}{*{20}{c}}
   {0} & {2 \kappa} & {} & {}  \\
   {0} & {0} & {} & {}  \\
   {} & {} & {\textbf{0}} & {}  \\
   {} & {} & {} & {\textbf{0}} \\
\end{array}} \right) \; \; .
\end{equation}

We have to determine the eigenvalues and the left and right eigenvectors of the matrix $\textbf{F}$:
\begin{subequations} \label{F_eigenthings}
\begin{gather}
\textbf{F} \; \textbf{r}^{(j)} \; = \; \omega_j \; \textbf{r}^{(j)} \\
\textbf{F}^\dag \; \textbf{l}^{(j)} \; = \; \omega_j^* \; \textbf{l}^{(j)} \; \; .
\end{gather}
\end{subequations}
These vectors form a bi-orthogonal basis, $\textbf{l}^{(k) \, \dag} \, \textbf{r}^{(j)} = \delta_{\, k \, j} \;$. The vector $\tildevec{v} (t)$ can be decomposed in terms of quasi-normal modes
\begin{equation} \label{normal_mode_exp}
\tildevec{v} (t) \; = \; \sum_j \; \tilde{\rho}_j (t) \; \textbf{r}^{(j)} 
\end{equation}
The equation of $\tilde{\rho}_k (t)$ reads 
\begin{equation} \label{normal_mode_eq}
\frac{d}{dt} \, \tilde{\rho}_k (t) \; = \; - i \, \omega_k \, \tilde{\rho}_k (t) \; + \hat{Q}_k (t) \; \; ,
\end{equation}
where $\hat{Q}_k (t)= \textbf{l}^{(k) \, \dag} \opvec{q} (t)$ is the projected noise. The solution is readily obtained
\begin{equation} \label{normal_mode_solution}
\tilde{\rho}_k (t) \, = \, \tilde{\rho}_k (0) \, e^{\, -i \, \omega_k \, t} \,
 + \, \int_0^t \, dt' \, \hat{Q}_k (t') \, e^{\, -i \, \omega_k \, (t - t')} \; .
\end{equation}

After some straightforward calculations, we get for the second order correlation functions in the steady state
\begin{equation} \label{second_order_correlations_result}
\ev{\hat{v}_\mu \, \hat{v}_\nu}_{\, st}= \sum_{n,\, j} \sum_{k, \, l} \, l^{(k) \dag}_n \, D_{n j} \, l^{(l) \dag}_j
 \frac{1}{ i (\omega_k + \omega_l )} \; r^{(k)}_\mu \, r^{(l)}_\nu \,.
\end{equation}
These are the second-order correlation functions needed for the HFB theory both for the mean field and for the fluctuation equations.  Let us emphasize at this point that the source of non-vanishing correlations is the quantum noise associated with the dissipative dynamics of the system. In this model the only dissipative process is the photon field decay, and correspondingly, the  correlation functions are proportional to $\kappa$.

\subsection{Iteration algorithm for the self-consistent solution}
\label{sec:Iteration}

In order to solve the coupled equations of the HFB theory self-consistently, we adopt an iteration algorithm. Here we  present it in some detail because the algorithm itself sheds more light on the structure of the cross-coupled mean-field and fluctuation equations. 

The fixed system parameters are  $N$, $\omega_R$, $\Delta_C$, $\eta_t$, $U_0$ and $\kappa$. Initially, the $\alpha$ and $\boldvec{\beta}$ mean-fields, and all the second-order correlation functions receive a random initial value.  The iteration algorithm is the following:
\begin{enumerate}
\item Determine the total number of condensed atoms $N_c$ from Eq.~\eqref{eq:N_BEC} and calculate then the coupling constants $y$ and $u$ from Eq.~\eqref{y,u_def}. 

\item Determine the matrix \textbf{M} from  Eq.~\eqref{M_def}. 

\item  Decouple the atomic modes: diagonalize the matrix \textbf{M} to get the orthogonal transformation $\textbf{O}$ according to Eq.~(\ref{matrix_orthogonal_transformation_a}), and then perform  the transformation of Eq.~(\ref{matrix_orthogonal_transformation_b}) to obtain the   matrices $\tildevec{M}^{(j)}$.

\item Calculate $\textbf{R}$ from \eqref{R_def} and evaluate the chemical potential $\mu$ from Eq.~\eqref{mu_def}. 

\item Update the value of $\boldvec{\beta}$ by using Eq.~\eqref{beta_mean_field}; note that the diagonal matrix $\left( \boldvec{\Lambda} -\mu \textbf{I} \right)$ is straightforwardly inverted. 

\item Update the value of the mean-field $\alpha$ from Eq.~\eqref{alfa_mean_field}  by using $\Omega$ from Eq.~\eqref{Omega_def_new_basis}.

\item Having the new mean field values,  the coefficients of the linearized differential equation~\eqref{eq:LinFluctNewBasis} can be inserted into the matrix $\textbf{F}$ of \eqref{lin_fluctuation_eq}, and the eigenvalues, the left- and right-eigenvectors are to be calculated \footnote{Since the condensate breaks the U(1) symmetry of the model, the Goldstone-theorem states that the spectrum of the system must contain a zero mode. This zero mode is $\tilde{b}_0$ and it describes the particle number and phase fluctuations of the condensate. 
But in the HFB approximation, the frequency of the $\tilde{b}_0$ mode is not strictly zero: this is a well-known weakness of the HFB approximation. This frequency is very small, but couples also to  $\tilde{a}$ and $\tilde{a}^\dag$, which renders the problem to be numerically unstable. To avoid it, we set to zero the frequency and the couplings related to the mode $\tilde{b}_0$ in the matrix $\textbf{F}$  by hand. The algorithm proved to be numerically stable.}.

\item Use Eq.~\eqref{second_order_correlations_result} to get the new second-order correlation functions. 

\item This step completes the iteration algorithm, return to step 1.

\end{enumerate}

The iteration is performed until it converges, which is tested, in our routine, by the variation of the solution for $\alpha$.  
At the end, we have the condensate wave function $\boldvec{\beta}$ and the second-order correlation functions of the modes $\tildevec{b}$. To obtain them in the original basis, $\boldvec{\gamma}$ and the correlation functions of the modes $\tildevec{c}$, one needs to perform the inverse of the  orthogonal transformation $\textbf{O}$, which has anyway been determined during the iteration.

\section{Finite-size scaling}
\label{sec:Results}

We showed that in the thermodynamic limit the open system exhibits critical behavior in the vicinity of the critical coupling \cite{Nagy2011Critical}
\begin{equation}
y_c=\sqrt{\omega_R \; \frac{\delta_C^2+\kappa^2}{\delta_C}} \;,
\end{equation}
where $\delta_C = -\Delta_C+ \tfrac{1}{2} N U_0$ is the shifted frequency of the cavity mode.
In the normal phase ($y<y_c$), the condensate wave function is homogeneous $\boldsymbol{\gamma} = (1,0,0,\ldots)$ and the mean cavity field is zero $\alpha_0 = 0$, while in the superradiant phase ($y>y_c$) a finite mean field builds up in the higher momentum modes of the condensate ($\gamma_i \neq 0$), and coherent light field appears in the cavity ($|\alpha_0| > 0$). On top of the mean fields, there are fluctuations strongly increasing near the critical point. The correlation functions of the fluctuations are singular and diverge in the Bogoliubov theory according to scaling laws.
This non-equilibrium phase transition is accompanied by critical slowing down, which is responsible for the scaling behavior. Quite remarkably the order parameter scales in a similar way as for the closed system. However, the susceptibilities characterizing the fluctuations are governed by completely different scaling laws. In this section we focus our attention mainly on the effects of finite particle number, that is finite-size scaling.

\subsection{Regularization of the criticality}

\begin{figure}[ht!]
\begin{center}
\includegraphics[width=0.8\columnwidth]{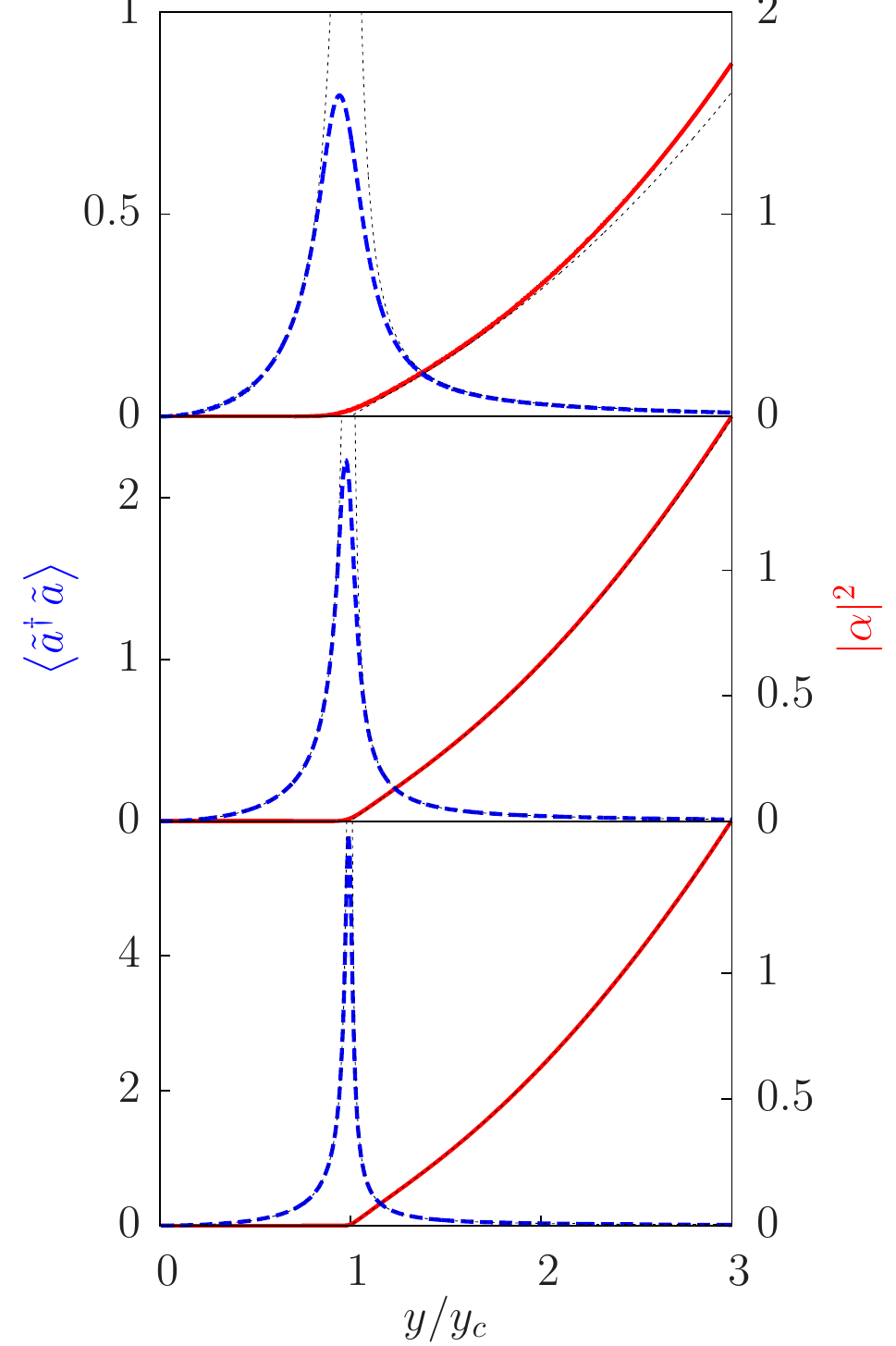}
\end{center}
\caption{(Color online) Coherent (solid line) and incoherent (dashed line) photon number in the cavity for an increasing number of atoms  $N=100$ (top), $N=1000$ (middle) and $N=10000$ (bottom). Thin dotted lines show the corresponding quantities in the thermodynamic limit ($N\rightarrow \infty$, calculated by multimode Bogoliubov theory). Parameters: $\omega_R=1$, $\Delta_C=-2$, $\kappa=2$, $U_0=0$ and $n_{\rm max}=2$. }
\label{y_alpha_abra}
\end{figure}

In the HFB theory, we couple the correlation functions of the fluctuations  back to the mean field equations. For finite atom number, this correction regularizes the critical divergence of the fluctuations, and account for nontrivial finite size effects and scaling. Fig.~\ref{y_alpha_abra} presents the cavity photon number and shows the convergence towards the singularity of the thermodynamic limit when the atom number is gradually increased. The photon number is split, according to the mean-field separation \eqref{a_c_mean_field_decomposition},
\begin{equation} \label{teljes_fotonszam_felbont}
\ev{\on{\hat{a}}} \; = \; N_c \; |\alpha|^2 \; + \; \ev{\on{\tilde{a}}} \; .
\end{equation}

It is convenient to plot separately the coherent and incoherent parts, since the $N_c$ factor gives rise to orders of magnitude differences. Moreover, the qualitative dependence is significantly different. The mean field reflects nicely that the non-analytic point in $y=y_c$ is smoothed out for finite $N$, however, the cross-over region becomes narrower as $N$ is increased. In parallel, the number of incoherent photons increases as a power law, $\mbox{max}\left\{\ev{\on{\tilde{a}}}\right\} \propto N^\tau$, with the exponent $\tau = 0.41\pm 0.02$. This result points out the sharp difference between the open and the closed Dicke models, since this exponent significantly differs from $\tfrac{1}{3}$, which is the exact finite-size exponent of the closed system \cite{Vidal2006Finitesize}. The bottom panel of the figure shows that the HFB results for $N=10000$ are practically indistinguishable from those of the  Bogoliubov theory (thin dotted lines) where the limit $N\rightarrow\infty$ is taken. For generating the dotted line results, we extended the model of Ref.~\cite{Nagy2011Critical} to include higher excited modes up to the same cutoff $n_{\rm max}=2$ \cite{Konya2011Multimode}.  

\begin{figure}[htbp]
\begin{center}
\includegraphics[width=0.8\columnwidth]{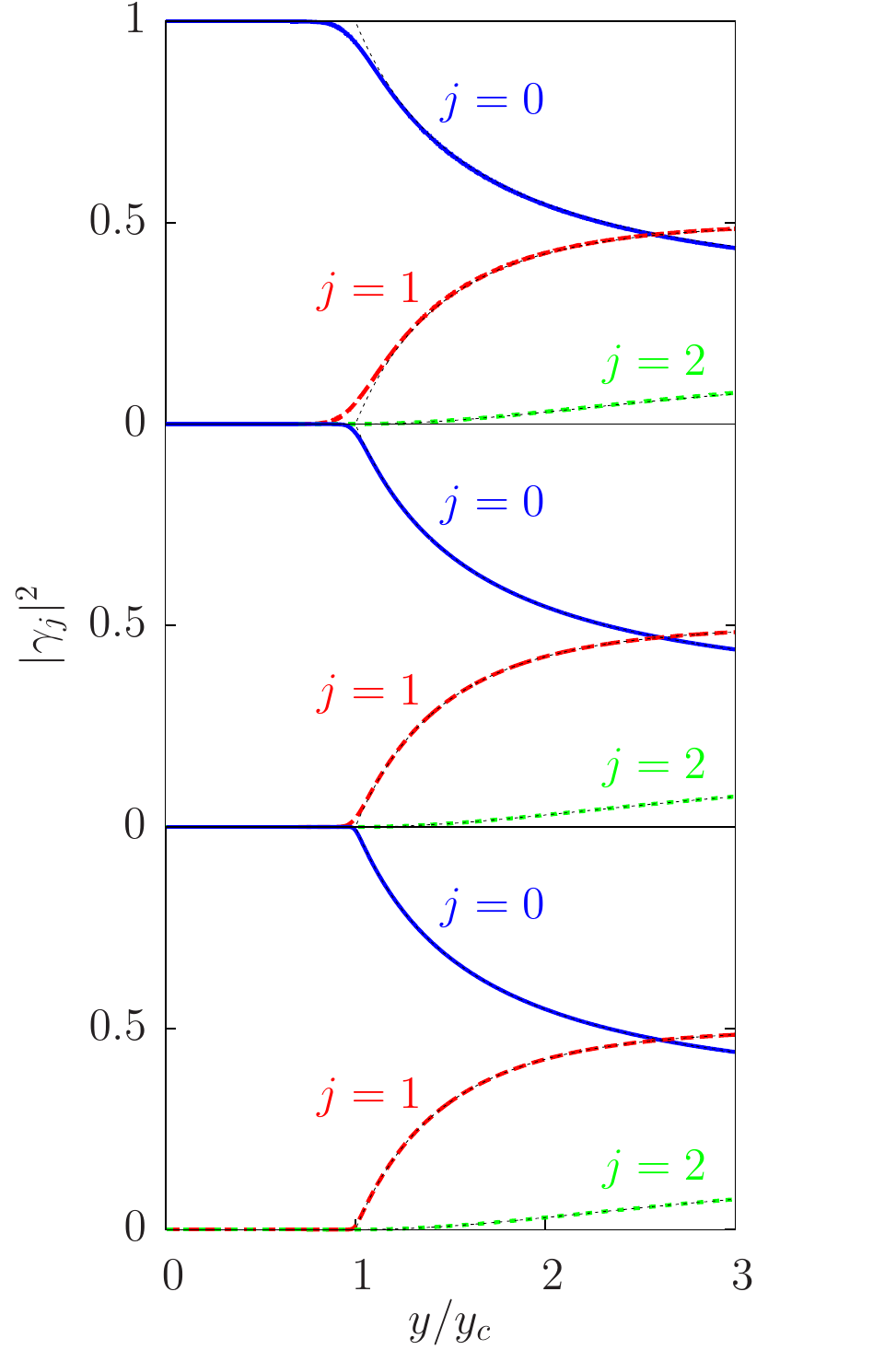}
\end{center}
\caption{(Color online) The condensate mean field components in the Fourier modes for increasing number of atoms $N=100$ (top), $N=1000$ (middle) and $N=10000$ (bottom). Thin dotted lines represent the results of the thermodynamic limit.
Parameters are the same as in Fig.~\ref{y_alpha_abra}.}
\label{y_gamma_abra}
\end{figure}
\begin{figure}[htbp]
\begin{center}
\includegraphics[width=0.8\columnwidth]{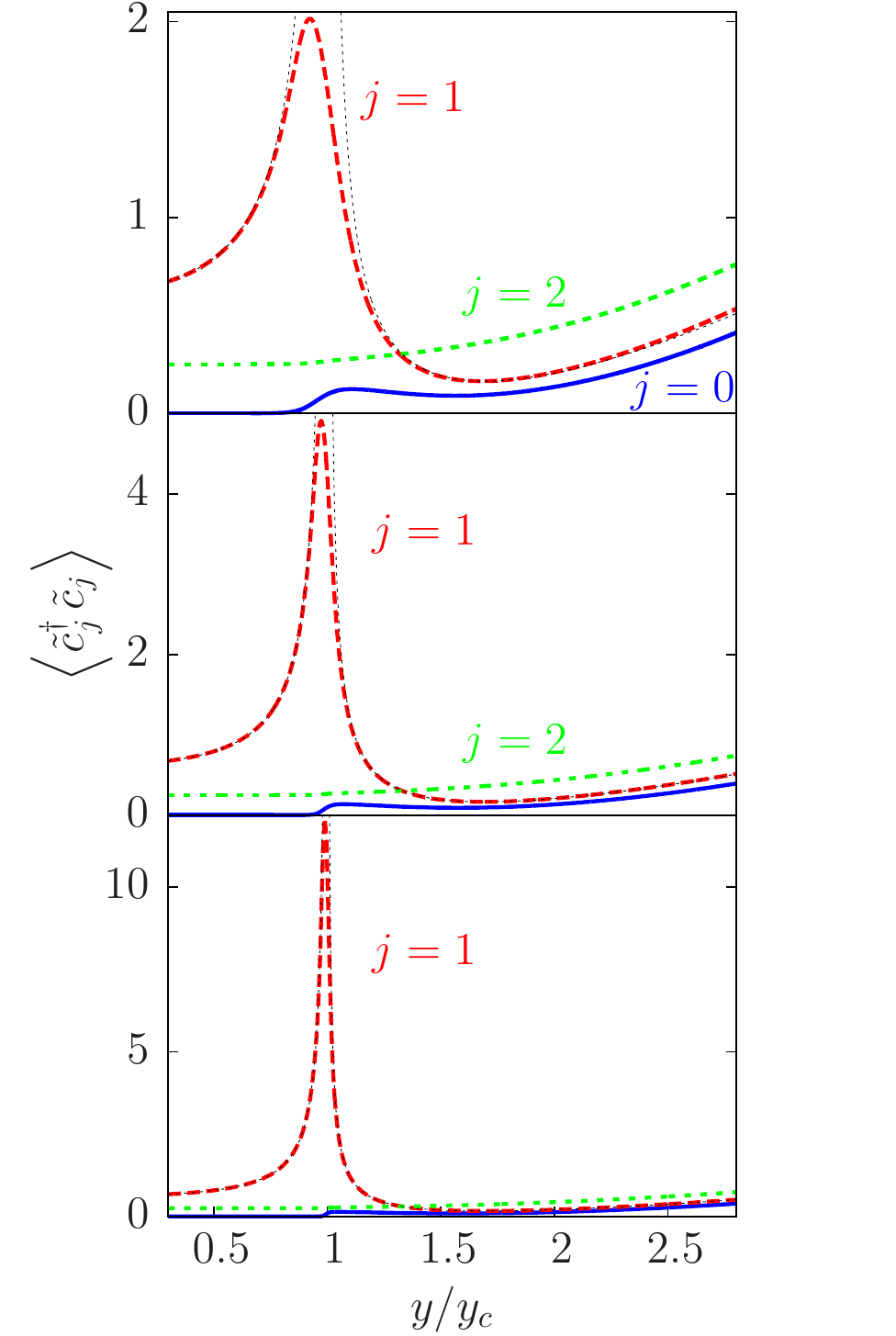}
\end{center}
\caption{(Color online) Incoherent excitation in the Fourier modes of the atomic motion above the condensate mean field for increasing number of atoms $N=100$ (top), $N=1000$ (middle) and $N=10000$ (bottom). Thin dotted lines represent the results of a multimode Bogoliubov theory of the thermodynamic limit. Parameters are the same as in Fig.~\ref{y_alpha_abra}.}
\label{y_cdagc_abra}
\end{figure}

The very same behavior is reflected in the populations of the atomic modes, shown in Fig.~\ref{y_gamma_abra} for the mean field and on Fig.~\ref{y_cdagc_abra} for the fluctuations. In addition to elucidating the formation of criticality, this plot reveals quantitatively the negligible effect of the higher excited  modes on the threshold region. The population $|\gamma_2|^2$ in the mode $\cos(2 k x)$ starts to increase only quadratically and only from the critical point. It follows then that the critical behavior of the coupled BEC-cavity system is equivalent to that of the Dicke model which corresponds to the two-mode approximation. 

\begin{figure}[htbp]
\begin{center}
\includegraphics[width=\columnwidth]{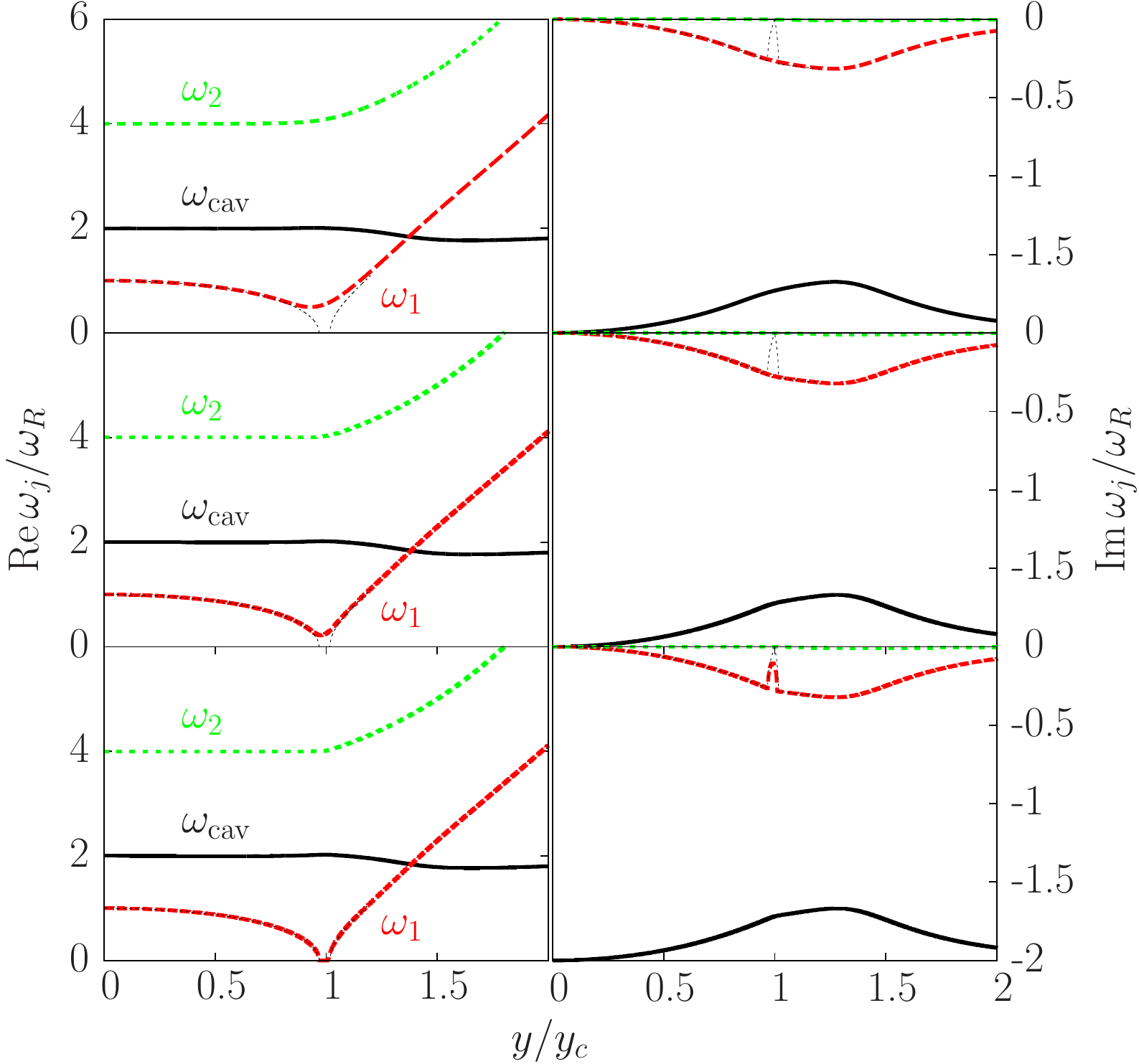}
\end{center}
\caption{(Color online) Spectrum of excitations: real part (left column) and imaginary part (right column) for $N=100$ (upper row), $N=1000$ (middle row) and $N=10000$ (bottom row). Thin dotted lines show the results of the thermodynamic limit.  Parameters are the same as in Fig.~\ref{y_alpha_abra}.}
\label{spektrum_abra}
\end{figure}
In Fig.~\ref{spektrum_abra} we present the spectrum of fluctuations for three different atom numbers. The spectrum is made of complex eigenfrequencies with non-vanishing, negative imaginary parts due to the cavity loss rate $\kappa$. By construction, it possesses the symmetry $\omega_j \leftrightarrow -\omega_j^{*}$, hence we plot only half of the spectrum with positive real parts. The spectra is labeled according to the $y=0$ value of the real part, that is, the curves starting from $\omega_R$, $-\Delta_C= 2 \omega_R$, and $4 \omega_R$ correspond to the mode $\tilde{c}_1$ ($\omega_1$), the cavity field mode $\tilde{a}$ ($\omega_{\rm cav}$), and the motional mode $\tilde{c}_2$ ($\omega_2$), respectively. These plots reveal how close the ``soft'' mode's frequency $\omega_1$ approaches zero near the critical point. In the thermodynamic limit (thin dotted lines), its real part is zero in a finite interval around $y_c$, that is a distinctive feature of the open-system phase transition \cite{Nagy2011Critical}. Besides this, the imaginary parts reflect that the motional modes mix with the lossy photon mode. Following the symmetry $\omega_1 \leftrightarrow \omega_1^{'} = -\omega_1^{*}$, at the two end points of the interval where the real parts of $\omega_1$, $\omega_1^{'}$ vanish, their imaginary parts bifurcates (only the upper branch is plotted). The critical point is reached when the upper branch hits zero. The finite-size scaling is manifested in the vanishing of the imaginary part, $\mbox{min.}\{|\im\,\omega_1|\} \propto N^{-\epsilon}$. We find the exponent $\epsilon = 0.44\pm 0.02$. Note that $\im\,\omega_1$ defines the correlation time of the system. The admixture of the mode  $\tilde{c}_2$ with the field mode is very small and the cavity decay affects this mode only marginally, what supports that the truncation of the Fourier set of modes, which is a basic ingredient in many papers, is well justified. 

Let us analyze the characteristic magnitude of finite-size corrections to the thermodynamic limit results. The finite-size effect is due to the back coupling of the second-order correlations of the modes  $\tilde{a}$ and $\tilde{c}_1$ into the mean field equations. It follows from \eqref{second_order_correlations_result} that the eigenfrequencies appear in the denominator, whereas $\kappa$ stands in the numerator from the diffusion matrix. The significant peaks of the correlation functions near the critical point, exhibited in Figs.~\ref{y_alpha_abra} and \ref{y_cdagc_abra}, are due to the soft mode frequency, $\omega_1$, that tends to vanish in the critical point. As the HFB terms in Eqs.~\eqref{eq:HFBmean} and \eqref{eq:HFBfluct} are multiplied by the condensate atom number $1/N_c \approx 1/N$, the order of magnitude characterizing the back coupling is
\begin{equation}
\frac{\kappa}{N \; \im\{\omega_1\}}\; .
\end{equation}
At the vicinity of the critical point, the back coupling scales with exponent $\epsilon-1 < 0$, thus the HFB corrections vanish for $N\rightarrow \infty$.  It follows then that the thermodynamic limit of the HFB theory renders the results of the Bogoliubov theory.

\subsection{Fano factor}
In the following we compare the HFB results to those of an exact numerical calculation with $n_{\rm max}=1$. This latter is performed by the Monte-Carlo Wave Function method using the general purpose C++QED package \cite{Vukics2007CQED,Vukics2012CQEDv2} and gives numerically exact solutions for the equations of motion in \eqref{eq:EquationsOfMotion}. We run the simulations for long enough time to reach the steady-state. 

The numerical method yields the mean photon number, for example, but it is not separated to the contributions of the mean field and that of the fluctuations. Therefore we need to find a quantity, other than the second-order noise correlations, which can characterize the criticality and can be assessed both in the HFB and numerically. This is the Fano factor
\begin{equation}
\label{eq:FanoFactor}
F= \frac{\ev{\left( \on{\hat{a}} \right)^2 } - \ev{\on{\hat{a}}}^2 } { \ev{\on{\hat{a}}} }  \;,
\end{equation}  
which contains the variance of the field amplitude, however, the quantum noise part is not suppressed by the mean field, this latter being magnified by the factor of $N_c$. For a coherent state the Fano-factor is 1 regardless of the amplitude. Deviation from unity thus reveals the increasing role of the fluctuations with respect to the mean, so that the appearance of a peak at the critical point can be expected. The scaling of the peak height with the atom number can be evaluated.

\begin{figure}[h!]
\begin{center}
\includegraphics[angle=0,width=0.75\columnwidth]{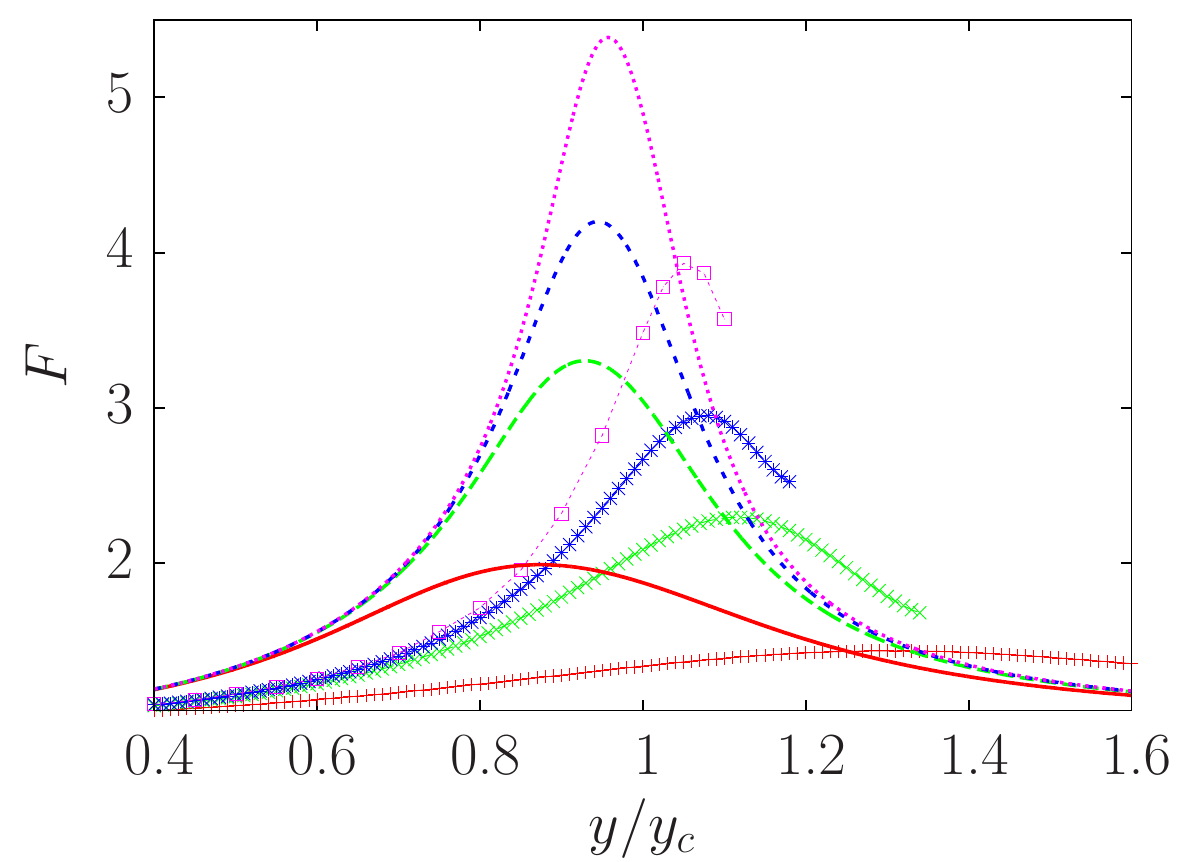}
\end{center}
\caption{(Color online) The Fano factor as a function of pump strength. Lines without line points show the HFB approximation, lines with line points correspond to the exact numerical results. The peaks in increasing order correspond to atom numbers $N=10$, $50$, $100$ and $200$. Parameters are the same as in Fig.~\ref{y_alpha_abra}. }
\label{y_Fano_abra}
\end{figure}

\begin{figure}[htb]
\begin{center}
\includegraphics[angle=0,width=0.75\columnwidth]{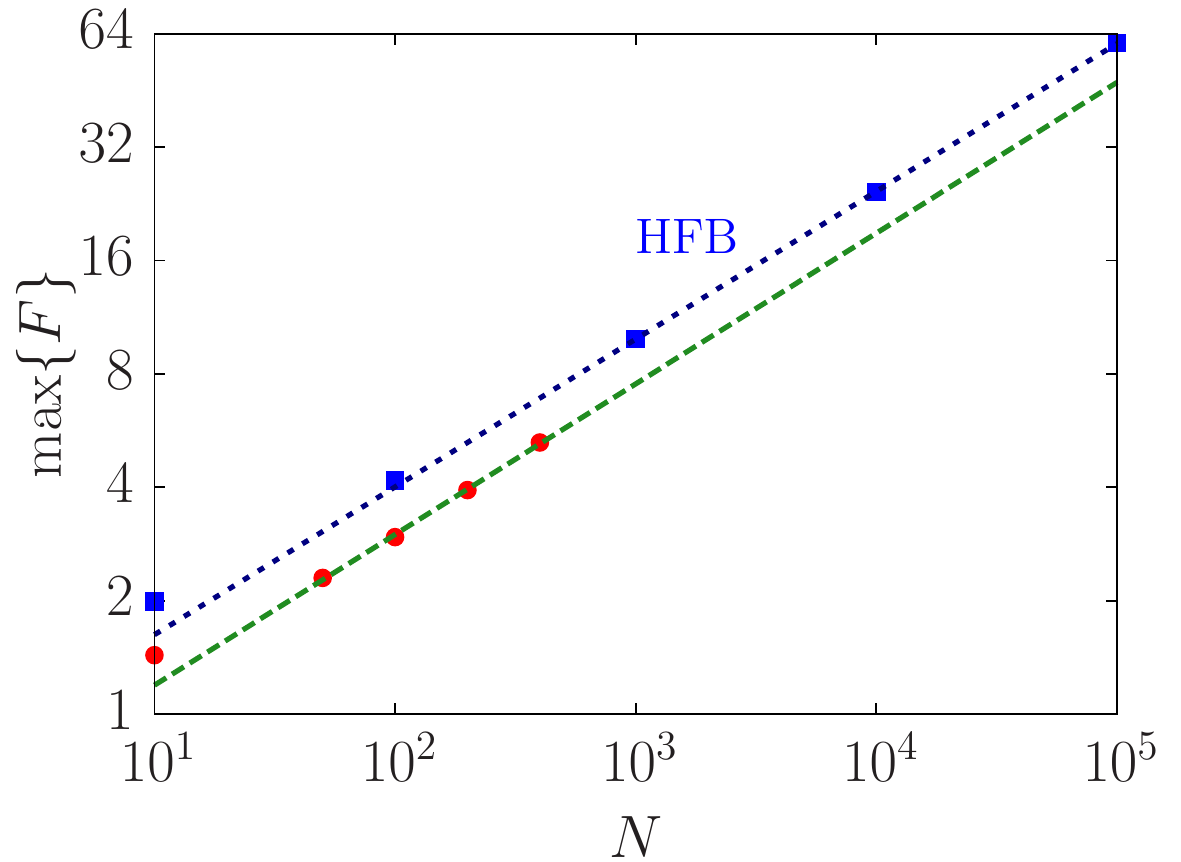}
\end{center}
\caption{(Color online) Critical exponents of the finite-size scaling of the peak height in the Fano-factor. Solid line shows a linear fit on the calculated points. The HFB calculation is performed for atom numbers  $N=10, \; 100, \; 1000, \; 10000, \; 100000$ (the fit discards the points $N=10$ and $N=100$), the numerical simulation is limited to 
$N=10, \; 50, \; 100, \; 200, \, 400$ (the fit is without $N=10$). Parameters are the same as in Fig.~\ref{y_alpha_abra}.}
\label{Fano_csucs_abra}
\end{figure}

The Fano factor within the HFB approximation is
\begin{equation} \label{Fano_szamolas}
F  = 1 + 2 \ev{\on{\tilde{a}}} + \frac{\alpha^{* \, 2} \ev{ \tilde{a} \tilde{a} } + \; \alpha^2 \ev{ \tilde{a}^\dag  \tilde{a}^\dag }}{|\alpha|^2 } + O \left( \frac{1}{N_c} \right) \,.
\end{equation}
This is compared to the results of the numerical calculation in Fig.~\ref{y_Fano_abra} as a function of the pump strength and for various atom numbers. In accordance with our expectation, well-behaved peaks appear for finite $N$. On increasing the atom number, the peaks get narrower and higher, indicating how the system tends to the singular behavior of the thermodynamic limit.  
Note that there is some difference in the height and the position of the corresponding peaks in the two approaches. The HFB leads to a peak position below, whereas the numerical simulation gives it above the critical point. The convergence properties are, however, amazingly similar. In Fig.~\ref{Fano_csucs_abra} the power law finite-size scaling of the peak height with the number of atoms is presented on a log-log scale. The exponent obtained from the fit on the HFB results (blue squares) $0.39\pm 0.02$ agrees well with the exponent acquired from the numerical approach (red circles) $0.40 \pm 0.01$.

\subsection{Scaling ansatz and finite-size exponents}

To discuss the results further and to achieve a deeper understanding let us first review shortly the scaling relations found in the Bogoliubov theory \cite{Nagy2011Critical}, i.e., for an infinite system. The critical behavior of the open-system Dicke model follows from the vanishing of the characteristic frequency $\omega_1$. Interestingly the real part vanishes in an extended interval of $y/y_c$ and the imaginary part that is responsible for the stability bifurcates. The critical point is reached when the upper branch of the imaginary part reaches zero. In the infinite system $\mathrm{Im}\ \omega_1\sim |y_c-y|^{\xi}$, with the critical exponent $\xi=1$.

From Eq. \eqref{second_order_correlations_result} one can deduce that the critical part of the second order correlation functions are dominated by the soft mode, that is $\langle v_\mu v_\nu \rangle\sim( \mathrm{Im}\ \omega_1)^{-1}\sim|y_c-y|^{-1}$. This critical behavior is in contrast to the case of the closed system Dicke model, where the soft mode only has a real part that vanishes as $|y_c-y|^{1/2}$ and consequently the second order correlations scale as $|y_c-y|^{-1/2}$.

To demonstrate the scaling properties to the finite size system we assume a scaling ansatz for the imaginary part of the soft mode frequency, namely we assume that it depends on a combination of the two variables $\tilde{y}\equiv(y_c-y)/y_c$ and $N$ rather than independently of the two. The most general form can be written as:
\begin{equation}
\label{scaling_ansatz}
\mathrm{Im}\ \omega_1\sim \tilde{y}^\xi\ \varphi(N^\epsilon \tilde{y}),
\end{equation}
with $\xi$ and $\epsilon$ are yet undetermined exponents and $\varphi(x)$ is a dimensionless function whose asymptotic values are fixed by the following argument. In the infinite system $N\rightarrow\infty$ and with $\tilde{y}$ fixed the argument $x\rightarrow\infty$ and $\mathrm{Im}\ \omega_1$ has to give back the scaling in the thermodynamic limit, i.e., $\xi=1$. Therefore $\lim_{x\rightarrow\infty}\varphi(x)=\text{const}$. Similarly for the finite system and exactly $\tilde{y}=0$ the frequency has to stay nonzero, therefore $\lim_{x\rightarrow0}\varphi(x)=x^{-1}$. Consequently the minimum value scales as $\mathrm{Im}\ \omega_1\sim N^{-\epsilon}$. For finite $N$ the location of the minimum value of $\mathrm{Im}\ \omega_1$ is also shifted from zero to  $\tilde{y}^\star$. Its scaling can be extracted by assuming that $\varphi(x)$ has a minimum value at a finite $x^\star$, therefore $\tilde{y}^\star=x^\star N^{-\epsilon}$.

The second order correlators $\langle \tilde{a}^\dagger \tilde{a}\rangle$, $\langle \tilde{c}_1^\dagger \tilde{c}_1\rangle$, and also $F$ can be separated to a nonuniversal finite and to a critical part. The critical part, denoted by $\Phi$, is proportional to $(\mathrm{Im}\ \omega_1)^{-1}$, as seen from Eq. \eqref{second_order_correlations_result}. Therefore all of these quantities scale according to
\begin{equation}
\Phi\sim\tilde{y} ^{-1}\Big[\varphi(N^\epsilon \tilde{y}) \Big]^{-1}.
\end{equation}
It follows then that for an infinite system all these correlation functions diverge around the critical point with the same exponent $\xi=1$. Moreover for a finite system all has the same finite size scaling exponent $\epsilon$ which can be read from the scaling of the heights and also by the location of the corresponding peaks. 

The scaling exponents of the HFB theory are collected in Table~\ref{exponents}.
It can be seen that $\epsilon\approx0.4$ in good agreement for all the correlation functions both from the heights and also from the locations of the peaks.

\begin{table}
\begin{tabular}{| c | c | c |}
\hline
Quantity & Prop. & Exp. \\ 
\hline\hline
$\langle\tilde{a}^\dag \tilde{a}\rangle$ & max. & $0.41$ \\ \cline{2-3}
  & $|y_{\rm max}-y_c|$ &  $\,\quad-0.39\quad\,$ \\ 
\hline
$\langle\tilde{c}_1^\dag\tilde{c}_1\rangle$ & max. & $0.39$ \\ \cline{2-3}
 &  $|y_{\rm max}-y_c|$ &  $-0.41$ \\
 \hline
$|\im\{\omega_1\}|$ & min. & $-0.44$ \\ \cline{2-3}
 & $|y_{\rm min}-y_c|$ & $-0.39$ \\
\hline
$F$ & max. & $0.39$ \\ \cline{2-3}
 & $|y_{\rm max}-y_c|$ & $-0.39$ \\
\hline
\end{tabular}
\caption{Summary of the finite-size exponents obtained from the HFB theory. A physical quantity $\Phi$ scale with exponent $\tau$ according to $\mbox{max}\{\Phi\} \propto N^\tau$. The estimated errors of the exponents are below $0.02$.}
\label{exponents}
\end{table}

\subsection{Atom-field entanglement}
\label{entanglement}

The cavity field mode and the atomic motional modes become entangled  in the steady state of the system. In the thermodynamic limit, the Bogoliubov approach leads to a non-vanishing finite logarithmic negativity even if the system is exposed to dissipation \cite{Nagy2011Critical}. Note that the ground state of the closed Dicke model is the two-mode squeezed state of which the logarithmic negativity diverges in the critical point \cite{Lambert2004Entanglement}. The damping in the open Dicke model regularizes then the divergence, however, the logarithmic negativity is still a non-analytic function of $y$ at the critical value $y_c$. In the following we present the finite-size corrections to the steady-state entanglement. 

The logarithmic negativity is an entanglement measure defined as \cite{Plenio2007Introduction}
\begin{equation}
E_{\cl{N}} \left( \tilde{\rho} \right)  =  \ln \left( \norm{\tilde{\rho}^{\; T_A}} \right) \; ,
\end{equation} 
where $\tilde{\rho}^{\; T_A}$ means the partial transpose of the density operator with respect to the subsystem $A$ and the trace norm is used, i.e., $\norm{O} = Tr\left\{ \sqrt{O^\dag  O} \right\}$. The bipartition, in our case, is such that subsystem $A$ is the photon mode and all the atomic motional modes form the complementary subsystem. In the HFB theory, the steady state is a Gaussian state for which the logarithmic negativity can be calculated from the symmetrically ordered correlation matrix \cite{Vidal2002Computable,Adesso2004Extremal},
\begin{equation} \label{correlation_matrix_def}
C_{j,k}= \frac{1}{2} \ev{\anticomm{\tilde{q}_j}{\tilde{q}_k}}\,,
\end{equation}
where $\anticomm{.}{.}$ denotes the anticommutator, and $\tilde{q}_j$ are the quadratures of the fluctuation operators, $\tilde{q}_0=\tilde{x} =\frac{1}{\sqrt{2}} \left( \tilde{a}^\dag   + \tilde{a} \right)$, $\tilde{q}_1=\tilde{p} = \frac{i}{\sqrt{2}} \left( \tilde{a}^\dag   - \tilde{a} \right)$, $\tilde{q}_{\, 2j}=\tilde{X}_j = \frac{1}{\sqrt{2}} \left( \tilde{b}^\dag_j + \tilde{b}_j \right)$, and $ \tilde{q}_{\, 2j+1}=\tilde{P}_j = \frac{i}{\sqrt{2}} \left( \tilde{b}^\dag_j - \tilde{b}_j \right)$. The elements of the correlation matrix $\textbf{C}$ are readily obtained from the second order correlation functions of Eq.~(\ref{second_order_correlations_result}) that have been derived in the course of the iteration algorithm. It can be easily shown that the correlation matrix of the partially transposed state $\textbf{C}^{T_A}$ is obtained by means of the substitution $\tilde{p} \rightarrow -\tilde{p}$ in the corresponding correlation matrix elements. The logarithmic negativity is then expressed by the symplectic eigenvalues $\tilde{\nu}_j$ of $\textbf{C}^{T_A}$ \cite{Vidal2002Computable} as
\begin{subequations}
\begin{gather}
E_{\cl{N}} \; = \; \sum_{j} \; f( \tilde{\nu}_j ) \\
f(\tilde{\nu}_j) = 
\begin{cases}
0 \; \; \; \; \; &\tilde{\nu}_j \ge \frac{1}{2} \\
-\mathrm{ln}\left( 2 \tilde{\nu}_j \right) \; \; \; \; \; &\tilde{\nu}_j < \frac{1}{2}
\end{cases} \; .
\end{gather}
\end{subequations}

\begin{figure}[htbp]
\begin{center}
\includegraphics[width=0.75\columnwidth]{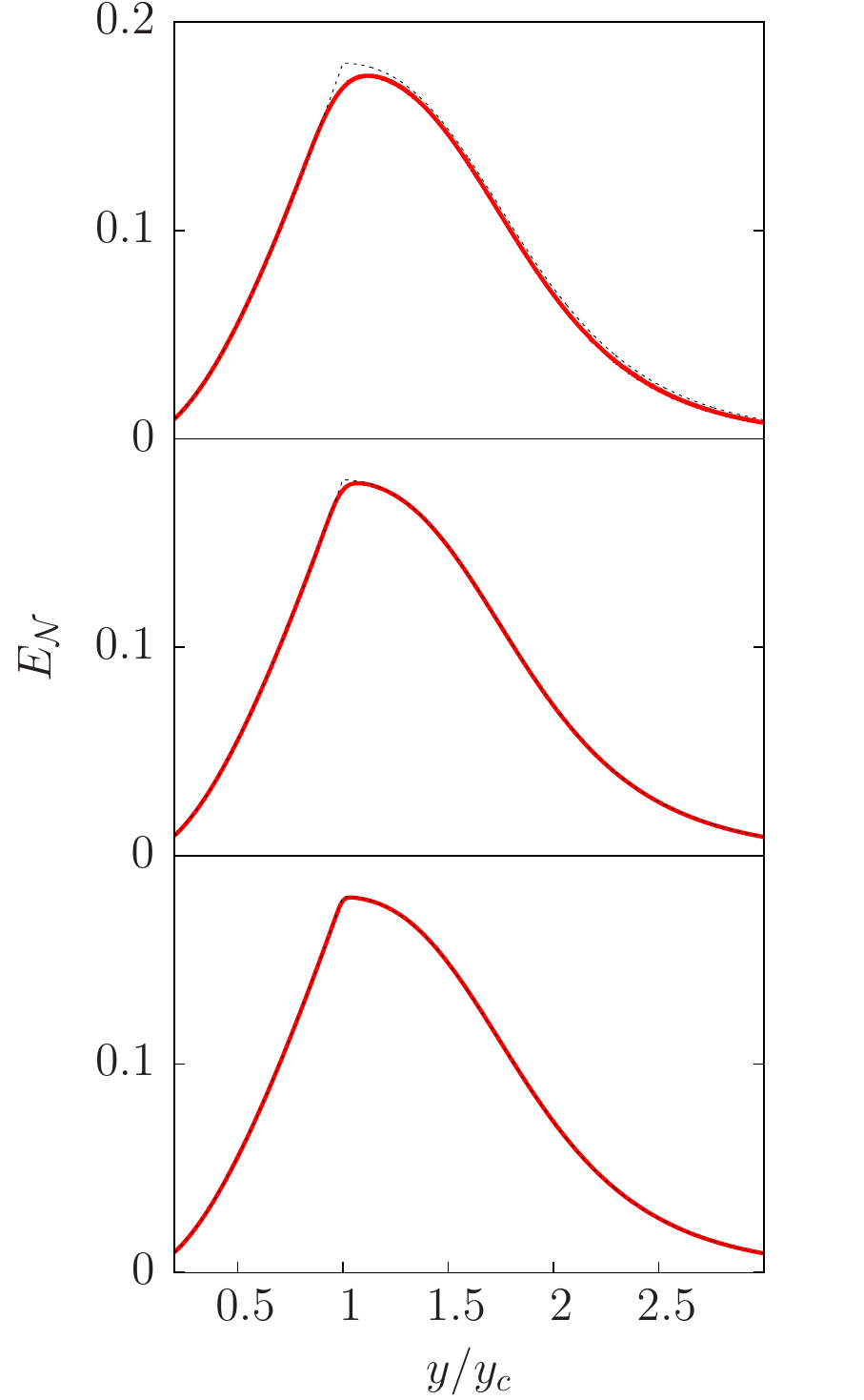}
\end{center}
\caption{The logarithmic negativity as a function of the pump strength. The atom numbers corresponding to the curves are $N=100$ (top), $N=1000$ (middle) and $N=10000$ (bottom). Parameters: $\omega_R=1$, $\Delta_C=-2$, $\kappa=2$, $U_0=0$ és $n_{\rm max}=2$. }
\label{y_EN_figure}
\end{figure}
We plot the logarithmic negativity as a function of the pump strength for three different atom numbers $N=100$,$1000$ and $10000$ in Fig.~\ref{y_EN_figure}. The atom--cavity entanglement is characterized by a broad peak with a maximum near the critical point, however, the peak height does not scale with the number of atoms $N$. It reaches the maximum of the thermodynamic limit for all finite $N$ shown.  On increasing $N$, the shape of the logarithmic negativity function  tends to reproduce the singular behavior with a jump of the derivative at $y=y_c$, found in the thermodynamic limit.

\section{Conclusion}
\label{sec:Conclusion}

Ultracold  atoms loaded into the small volume of a high-finesse resonator allow for studying quantum many-body physics and correlated state of matter originating from a peculiar, long-range interaction between the particles. In this paper we made an important step forward in the exploration of the self-organization process of atoms in a cavity, which is a quantum phase transition manifesting clearly the collective behavior of a quantum gas.  Going beyond the leading-order Bogoliubov-type description, here we developed the Hartree-Fock theory to treat the long-range interaction, and in particular, the steady-state of the damped-driven open system.  As a main merit of the theory, it renders the finite-size corrections to the bare mean-field theory, regularizes the singularity in the critical point. Finally, we were able to derive the finite-size scaling of the of the divergent correlation functions. The power law has been found together with the critical exponent which is in a very good agreement with the outcome of an exact numerical simulation.

\section{Acknowledgements}

This work was supported by the Hungarian National Office for Research and Technology under the contract ERC\_HU\_09 OPTOMECH, the Hungarian National Research Fund (OTKA T077629) and the Hungarian Academy of Sciences  (Lend\"ulet Program, LP2011-016).


\appendix

\begin{widetext}

\section{Heisenberg-equations after the mean-field decomposition}
\label{Heisenberg_mean-field_dec}

For completeness, the general equations of motion \eqref{eq:EquationsOfMotion}  are written out explicitly here by using the mean-field decomposition \eqref{a_c_mean_field_decomposition}. The terms are grouped according to the different powers of the fluctuation operators involved.  The equation of motion for the cavity field mode is
\begin{multline} \label{alpha_tilde{a}_eq_of_motion}
\sqrt{N_c} \; i \frac{d}{dt} \alpha (t) + i \frac{d}{dt} \tilde{a} (t) = 
\sqrt{N_c} {\Biggl[} \left( -\Delta_c + u \, \boldvec{\gamma}^\dag \textbf{M}^{(2)} \boldvec{\gamma} - i \kappa \right) \alpha + \frac{1}{2} y \, \boldvec{\gamma}^\dag \textbf{M}^{(1)} \boldvec{\gamma} {\Biggr]} \\
+ {\Biggl[} \left( -\Delta_c + u \, \boldvec{\gamma}^\dag \textbf{M}^{(2)} \boldvec{\gamma} - i \kappa \right) \tilde{a} + u \, \alpha \left( \boldvec{\gamma}^\dag \textbf{M}^{(2)} \tildevec{c} +
\tildevec{c}^\dag \textbf{M}^{(2)} \boldvec{\gamma} \right) + \frac{1}{2} y \, \left( \boldvec{\gamma}^\dag \textbf{M}^{(1)} \tildevec{c} + \tildevec{c}^\dag \textbf{M}^{(1)} \boldvec{\gamma} \right)
+ i \hat{\xi} (t) {\Biggr]} \\
+\frac{1}{\sqrt{N_c}} {\Biggr[} u \; \tildevec{c}^\dag \textbf{M}^{(2)} \tildevec{c} \; \alpha +
u  \left( \tildevec{c}^\dag \textbf{M}^{(2)} \boldvec{\gamma} + \boldvec{\gamma}^\dag \textbf{M}^{(2)} \tildevec{c} \right) \tilde{a}
+ \frac{1}{2} \, y \; \tildevec{c}^\dag \textbf{M}^{(1)} \tildevec{c} {\Biggr]} \\
+\frac{1}{N_c} {\Biggl[} u \; \tildevec{c}^\dag \textbf{M}^{(2)} \tildevec{c} \; \tilde{a} {\Biggr]} \; ,
\end{multline}
whereas for the matter wave field modes
\begin{multline} \label{gamma_tilde{c}_eq_of_motion}
\sqrt{N}_c \; i \frac{d}{dt} \boldvec{\gamma} + i \frac{d}{dt} \tildevec{c}  
=\sqrt{N_c} {\Biggl[} \left( \omega_R \; \textbf{M}^{(0)} + \frac{1}{2} y \left( \alpha^* +\alpha \right) \textbf{M}^{(1)} + u \alpha^* \alpha \, \textbf{M}^{(2)} -\mu \textbf{I} \right) \boldvec{\gamma}{\Biggr]} \\
+{\Biggl[} \left( \omega_R \textbf{M}^{(0)} + \frac{1}{2} y \left( \alpha^* +\alpha \right) \textbf{M}^{(1)} + u \alpha^* \alpha \, \textbf{M}^{(2)} - \mu \textbf{I} \right) \tildevec{c} 
+\left( \; \frac{1}{2} y \left( \tilde{a}^\dag + \tilde{a} \right) \textbf{M}^{(1)}
+ u \left( \alpha^* \, \tilde{a} + \tilde{a}^\dag \, \alpha \right) \textbf{M}^{(2)} \right) \boldvec{\gamma} {\Biggr]} \\
+\frac{1}{\sqrt{N_c}} {\Biggl[}  \left( \; \frac{1}{2} y \, \left( \tilde{a}^\dag + \tilde{a} \right) \, \textbf{M}^{(1)}
 + u \, \left( \alpha^* \, \tilde{a} + \tilde{a}^\dag \, \alpha \right) \textbf{M}^{(2)} \right) \tildevec{c}
 + u \; \on{\tilde{a}} \, \textbf{M}^{(2)} \boldvec{\gamma}  {\Biggr]} \\
+\frac{1}{N_c}{\Biggl[} u \; \on{\tilde{a}} \; \textbf{M}^{(2)} \tildevec{c} {\Biggr]} \; .
\end{multline}
\end{widetext}
The mean field equations are obtained exactly by taking the quantum average. Then, by subtracting the mean field equations form the above ones, the equations of motion for the fluctuations $\tilde{a}$ and $\tildevec{c}$  can be deduced.

The Hartree-Fock-Bogoliubov method relies on the following approximations, in \eqref{alpha_tilde{a}_eq_of_motion},
\begin{subequations}
\begin{gather*}
\tildevec{c}^\dag \textbf{M}^{(j)} \tildevec{c} \; \approx \; \ev{ \tildevec{c}^\dag \textbf{M}^{(j)} \tildevec{c} } \\
\left( \tildevec{c}^\dag \textbf{M}^{(2)} \boldvec{\gamma} + \boldvec{\gamma}^\dag \textbf{M}^{(2)} \tildevec{c} \right) \tilde{a}
\; \approx \; \ev{\tildevec{c}^\dag \, \tilde{a}} \textbf{M}^{(2)} \boldvec{\gamma} 
+ \boldvec{\gamma}^\dag \textbf{M}^{(2)} \ev{\tildevec{c} \, \tilde{a}} \\
\tildevec{c}^\dag \textbf{M}^{(2)} \tildevec{c} \; \tilde{a} \; \approx \; \ev{ \tildevec{c}^\dag \textbf{M}^{(2)} \tildevec{c} } \; \tilde{a}
 + \ev{\tildevec{c}^\dag \, \tilde{a}} \textbf{M}^{(2)} \tildevec{c} + \tildevec{c}^\dag \textbf{M}^{(2)} \ev{\tildevec{c} \, \tilde{a}}
\end{gather*}
\end{subequations}
and in \eqref{gamma_tilde{c}_eq_of_motion},
\begin{subequations}
\begin{gather*}
\left( \tilde{a}^\dag +\tilde{a} \right) \textbf{M}^{(1)} \tildevec{c} \; \approx \; \textbf{M}^{(1)} \left( \ev{\tilde{a}^\dag \tildevec{c}}
+ \ev{\tilde{a} \tildevec{c}} \right) \\
\left( \alpha^* \, \tilde{a} + \tilde{a}^\dag \, \alpha \right) \textbf{M}^{(2)} \, \tildevec{c} 
\; \approx \; \alpha^* \; \textbf{M}^{(2)} \ev{ \tilde{a} \tildevec{c} } + \alpha \; \textbf{M}^{(2)} \ev{\tilde{a}^\dag \tildevec{c}} \\
\on{\tilde{a}} \; \approx \; \ev{\on{\tilde{a}}} \\
\on{\tilde{a}} \textbf{M}^{(2)} \tildevec{c}  \approx  \ev{\on{\tilde{a}}}  \textbf{M}^{(2)} \tildevec{c} + \tilde{a}^\dag  \textbf{M}^{(2)} \ev{\tilde{a} \, \tildevec{c}} + \textbf{M}^{(2)} \ev{\tilde{a}^\dag \, \tildevec{c}} \tilde{a}
\end{gather*}
\end{subequations}
These substitutions lead to the model presented in Sec.~\ref{sec:HartreeFockBogoliubov}. 

To be specific, the first and third rows of the right-hand sides give the mean field equation \eqref{eq:HFBmean}, and the second and fourth rows give the fluctuation dynamics in Eq.~\eqref{eq:HFBfluct}.


\end{document}